\documentclass[9pt,twocolumn,twoside]{pnas-new} 

\templatetype{pnasresearcharticle}
\setboolean{displaywatermark}{false}

\usepackage{amsmath,amsfonts,amssymb}
\usepackage{graphicx}

\title{Estimation of blood oxygenation with learned spectral decoloring for quantitative photoacoustic imaging (LSD-qPAI)}

\leadauthor{Gröhl}

\author[ a,b]{Janek Gr\"ohl}
\author[ a,c]{Thomas Kirchner}
\author[ a,d]{Tim Adler}
\author[ a,b]{Lena Maier-Hein}
\affil[a]{Division of Computer Assisted Medical Interventions (CAMI), German Cancer Research Center (DKFZ), 69120 Heidelberg, Germany}
\affil[b]{Medical Faculty, Heidelberg University, 69120 Heidelberg, Germany}
\affil[c]{Faculty of Physics and Astronomy, Heidelberg University, 69120 Heidelberg, Germany}
\affil[d]{Faculty of Mathematics and Computer Science, Heidelberg University, 69120 Heidelberg, Germany}

\correspondingauthor{Correspondence: j.groehl@dkfz-heidelberg.de}


\begin{abstract}
One of the main applications of photoacoustic (PA) imaging  is the recovery of functional tissue properties, such as blood oxygenation (sO$_2$). This is typically achieved by linear spectral unmixing of relevant chromophores from multispectral photoacoustic images. Despite the progress that has been made towards quantitative PA imaging (qPAI), most sO$_2$ estimation methods yield poor results in realistic settings. In this work, we tackle the challenge by employing learned spectral decoloring for quantitative photoacoustic imaging (LSD-qPAI) to obtain quantitative estimates for blood oxygenation. LSD-qPAI computes sO$_2$ directly from pixel-wise initial pressure spectra $S_{\textrm{p}_0}$, which are vectors comprised of the initial pressure at the same spatial location over all recorded wavelengths. Initial results suggest that LSD-qPAI is able to obtain accurate sO$_2$ estimates directly from multispectral photoacoustic measurements \emph{in silico} and plausible estimates \emph{in vivo}.
\end{abstract}
 
\begin{document} 
\maketitle
\thispagestyle{fancy}
\ifthenelse{\boolean{shortarticle}}{\ifthenelse{\boolean{singlecolumn}}{\abscontentformatted}{\abscontent}}{}

\section{Introduction}
\label{sec:intro}

Photoacoustic (PA) imaging is a medical imaging modality that offers an optical signal contrast up to several centimeters deep inside tissue. In the last two decades, a lot of progress has been made towards its translation into clinical routine. However, the accurate and robust quantification of optical tissue properties or derived functional tissue properties still remains a major challenge~\cite{cox2009challenges,wang2012photoacoustic}. 
In order to obtain functional tissue properties such as blood oxygen saturation (sO$_2$), spectral unmixing algorithms are used to decompose a multispectral signal and to determine the quantity of specific chromophores that contributed to the signal spectrum over the wavelengths. For this process, the core assumption is that the acquired PA signal image $I$ - which is ideally an approximate reconstruction from the initial pressure distribution $\textrm{p}_0$ - is proportional to the optical absorption coefficient ($I \approx \textrm{p}_0 \propto \mu_a$) \cite{li2018photoacoustic}. However, this assumption does not always hold because $\textrm{p}_0$ is mainly proportional to the product of optical absorption $\mu_a$ and the light fluence $\phi$ ($I \approx \textrm{p}_0 \propto \mu_a \cdot \phi$). Hence, the fluence has an influence on the recorded spectra and potentially leads to large errors in sO$_2$ quantification.

To overcome this issue, part of the research in the field of quantitative PA imaging (qPAI) aims at compensating for these fluence effects (cf. e.g.~\cite{maslov2007effects,bu2012model,zhao2017optical,vogt2019photoacoustic}). For a long time, the field has focused on model-based approaches to extract quantitative information on optical tissue properties \cite{cox2006two,laufer2006quantitative} which typically suffer from long computation times. More recently, also machine learning-based approaches to qPAI have been published~\cite{kirchner2018context,cai2018end,grohl2018confidence}. These methods are often substantially faster than model-based algorithms but have not yet been demonstrated to work accurately and robustly in realistic \emph{in vitro} settings or \emph{in vivo}. Due to the lack of fast and accurate fluence compensation algorithms, most researchers and applications default to the simple linear unmixing algorithms in order to provide qualitative rather than quantitative values for relevant functional tissue parameters. 

This work tackles quantification of the functional tissue property sO$_2$ by introducing learned spectral decoloring for quantitative photoacoustic imaging (LSD-qPAI). LSD-qPAI is based on the assumption that there is a substantial benefit in considering fluence effects when quantifying sO$_2$~\cite{vogt2019photoacoustic,tzoumas2016eigenspectra,brochu2017towards}, compared to spectrally unmixing hemoglobin (Hb) and oxyhemoglobin (HbO$_2$) with commonly used linear methods~\cite{li2018photoacoustic} that neglect the aforementioned fluence effects. Previous deep learning approaches to qPAI try to estimate optical absorption from PA measurements and then derive functional tissue properties from these estimations, potentially leading to error propagation~\cite{kirchner2018context}. In contrast to this, we propose to directly estimate the functional tissue parameter sO$_2$ from pixel-wise $\textrm{p}_0$ spectra $S_{\textrm{p}_0}$, which is a vector of the initial pressure at the same spatial location over all recorded wavelengths. For this, we use multispectral \emph{in silico} $\textrm{p}_0$ training data, as illustrated in Figure \ref{fig:method}. This way, we force the machine learning algorithm to account for fluence effects in the $\textrm{p}_0$ spectra during sO$_2$ estimation. According to initial results, our \emph{in vivo} sO$_2$ estimates are physiologically more plausible when compared to linear spectral unmixing techniques.

\section{Materials and Methods}
\label{sec:materials}

\subsection{Concept overview}

\begin{figure} [h!t]
\begin{center}
\begin{tabular}{c} 
\includegraphics[width=5cm]{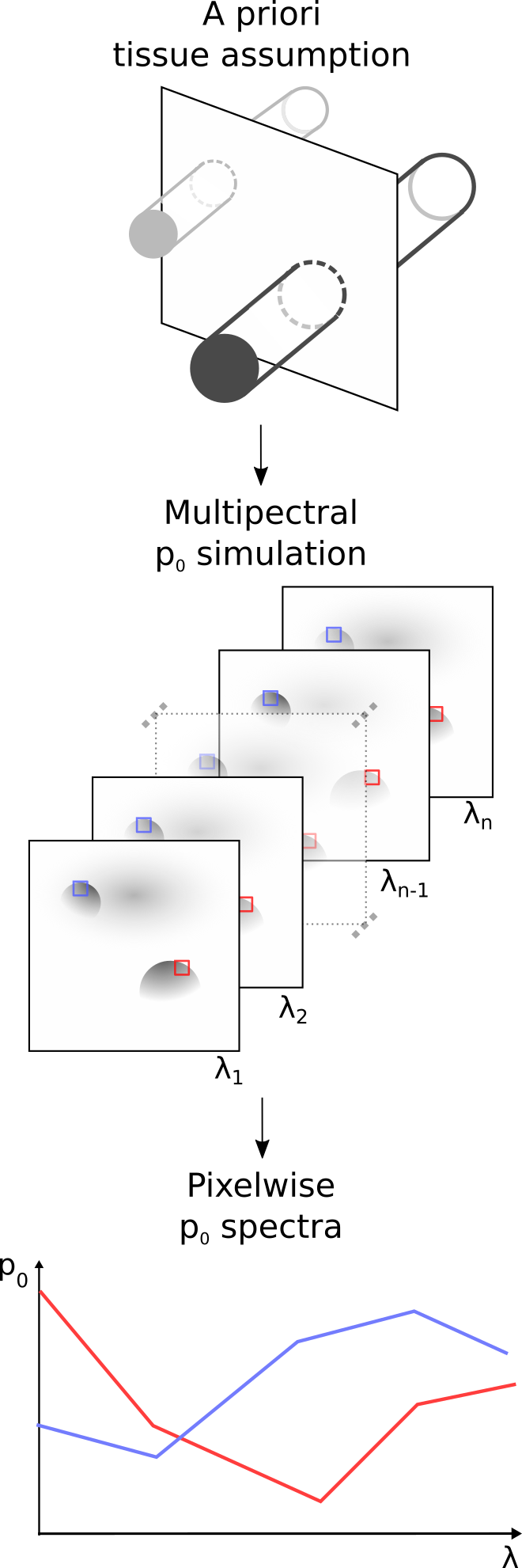}
\end{tabular}
\end{center}
\caption[example] 
{ \label{fig:method} 
Overview of the generation of $\textrm{p}_0$ spectra. An  \emph{in silico} tissue phantom with a ground truth tissue model is simulated over multiple wavelengths. The wavelength dependent behavior of the $\textrm{p}_0$ values in each pixel then defines the $\textrm{p}_0$ spectra.}
\end{figure}

With LSD-qPAI we approximate a function $f$ to estimate sO$_2$ from initial pressure $\textrm{p}_0$ spectra $S_{\textrm{p}_0}$ ($f\colon S_{\textrm{p}_0} \in \mathbb{R}^n \rightarrow \textrm{sO}_2 \in \mathbb{R}$). It is a data-driven method in which a neural network learns to compensate for different extents of \textit{spectral coloring} within a given $\textrm{p}_0$ spectrum, where the term spectral coloring refers to changes in the spectrum at a given spatial location due to wavelength-dependent absorption in the surrounding tissue \cite{tzoumas2014unmixing}. To account for the lack of real data comprising reliable reference or even ground truth ${sO}_2$ values given $\textrm{p}_0$ recordings, we create a dataset of pixel-wise $\textrm{p}_0$ spectra with various degrees of spectral coloring obtained from Monte Carlo simulated \emph{in silico} multispectral $\textrm{p}_0$ images (cf. Figure \ref{fig:method}). The variety of possibilities for spectral coloring is simulated by extracting $\textrm{p}_0$ spectra from different spatial locations within the image (e.g. shallow or deep background structures and superficial of deep vascular structures). With the generation of a representative dataset, the network learns to account for spectral coloring when inferring $\textrm{sO}_2$ from input $\textrm{p}_0$ spectra.

During training, the algorithm is presented tuples $(S_{\textrm{p}_0},\,\textrm{sO}_2)$, with ${S_{\textrm{p}_0}} \in \mathbb{R}^n$ and sO$_2 \in \mathbb{R}$. Here, each spectrum is normalized such that $\sum_{i}^{n} {\textrm{p}_0}_{\lambda_i} = 1$. When estimating oxygenation for recorded \emph{in vivo} spectra these are normalized as well - sacrificing $\textrm{p}_0$ amplitude information to eliminate the need to calibrate the \emph{in silico} $\textrm{p}_0$ training data to the target domain.

\subsection{Prototype implementation}

In the following paragraphs, we provide detailed descriptions of the prototype implementation of our approach, namely the simulation pipeline, the used deep learning model, and a linear spectral unmixing method.

\subsection*{Simulation pipeline}

We create simple homogeneous tissue volumes comprising tubular vessel structures that run orthogonal to the imaging plane. We simulated the light transport in this medium with a Monte Carlo method and simulated each of the \emph{in silico} vessel phantoms with 26 wavelengths equidistant from 700\,nm to 950\,nm in 10\,nm steps. For the Monte Carlo simulation, we used a multithreaded adaptation of the Monte Carlo framework \emph{mcxyz} \cite{jacques2014coupling} with $10^7$ photons for each simulation on a 0.6\,mm grid.

\subsection*{Deep learning model}

\begin{figure} [ht]
\begin{center}
\begin{tabular}{c} 
\includegraphics[width=8cm]{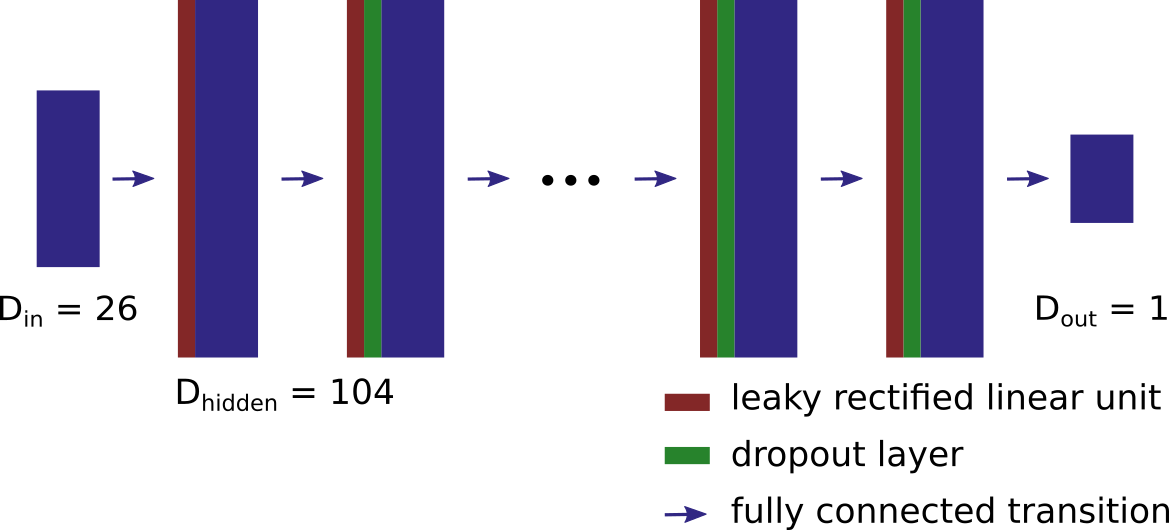}
\end{tabular}
\end{center}
\caption[example] 
{ \label{fig:network} 
Deep learning architecture for blood oxygenation (sO$_2$) reconstruction, comprising leaky rectified linear units (red), dropout layers (green), and layers of neurons (blue). In this setup, each of the neural layers is fully connected to the previous layer.}
\end{figure}

For reconstruction of sO$_2$ from $S_{\textrm{p}_0}$, we use a simple fully connected neural network architecture (cf. Figure \ref{fig:network}). We implement the network with eight hidden layers with four times the size of the input layer using \emph{pytorch} \cite{paszke2017automatic} and perform the experiments with the \emph{trixi} framework \cite{zimmerer_david_mic_dkfz_trixi_2018}. We use an $L$1 loss function, a learning rate of 10$^{-4}$, a batch size of 2000, with 1000 batches per epoch, and train the network for 25 epochs. The size of the input layers corresponds to the number of simulated or measured wavelengths (26 in this study) and we have a one dimensional output layer, which corresponds to the sO$_2$ estimation as the target parameter.

\subsection*{Linear spectral unmixing}

We use a linear spectral unmixing algorithm as a reference for the \emph{in vivo} imaging results. Specifically, we use the C++ \emph{Eigen} \cite{eigenweb} implementation of the QR householding matrix decomposition to unmix the recorded spectra for Hb and HbO$_2$ and calculate sO$_2$ from the ratio of these. 

\section{Experiments and Results}
\label{sec:results}

In the following section we outline the experiments we conducted for validation of the presented approach. The purpose of our experiments was two-fold: we first show the feasibility of our method with an \emph{in silico} evaluation on a held-out set from the same distribution as the training data. Secondly, we conduct feasibility tests on two different \emph{in vivo} datasets, of which the first contains PA images of an open porcine brain and the second dataset comprises PA images of the human forearm. These datasets are chosen because the compositions of relevant chromophores in tissue are very different in both cases. For both of these \emph{in vivo} settings we also compute sO$_2$ with a linear spectral unmixing algorithm to provide a reference.

\subsection*{Simulation parameters}

The key to the general applicability of the LSD-qPAI method is the simulation of a representative dataset of $\textrm{p}_0$ spectra. For this initial study, we created a total of 971 \emph{in silico} phantoms containing 2\,-\,10 tubular vessel-mimicking structures in a homogeneous background at random locations orthogonal to the imaging plane. For tissue mimicking \emph{a priori} conditions, we assumed blood vessels to have a hemoglobin concentration of 150 g/dl \cite{jacques2013optical}, generic background tissue to have a blood volume fraction of 2\%, and also consider an average of 5\% fat and 80\% water in this background tissue. The tubular vessel structures have a radius randomly drawn from a uniform distribution between 0.5\,mm and 4\,mm. Each vessel phantom was assigned a distinct constant blood oxygenation in vessel structures and the same constant blood oxygenation in the background. The tissue absorption coefficient was calculated for each voxel based on these assumptions and we set the reduced scattering coefficient to be constant over all wavelengths with $\mu'_s = $15\,cm$^{-1}$.

\subsection*{In silico validation}

\begin{figure} [ht]
\begin{center}
\begin{tabular}{c} 
\includegraphics[width=.45\textwidth]{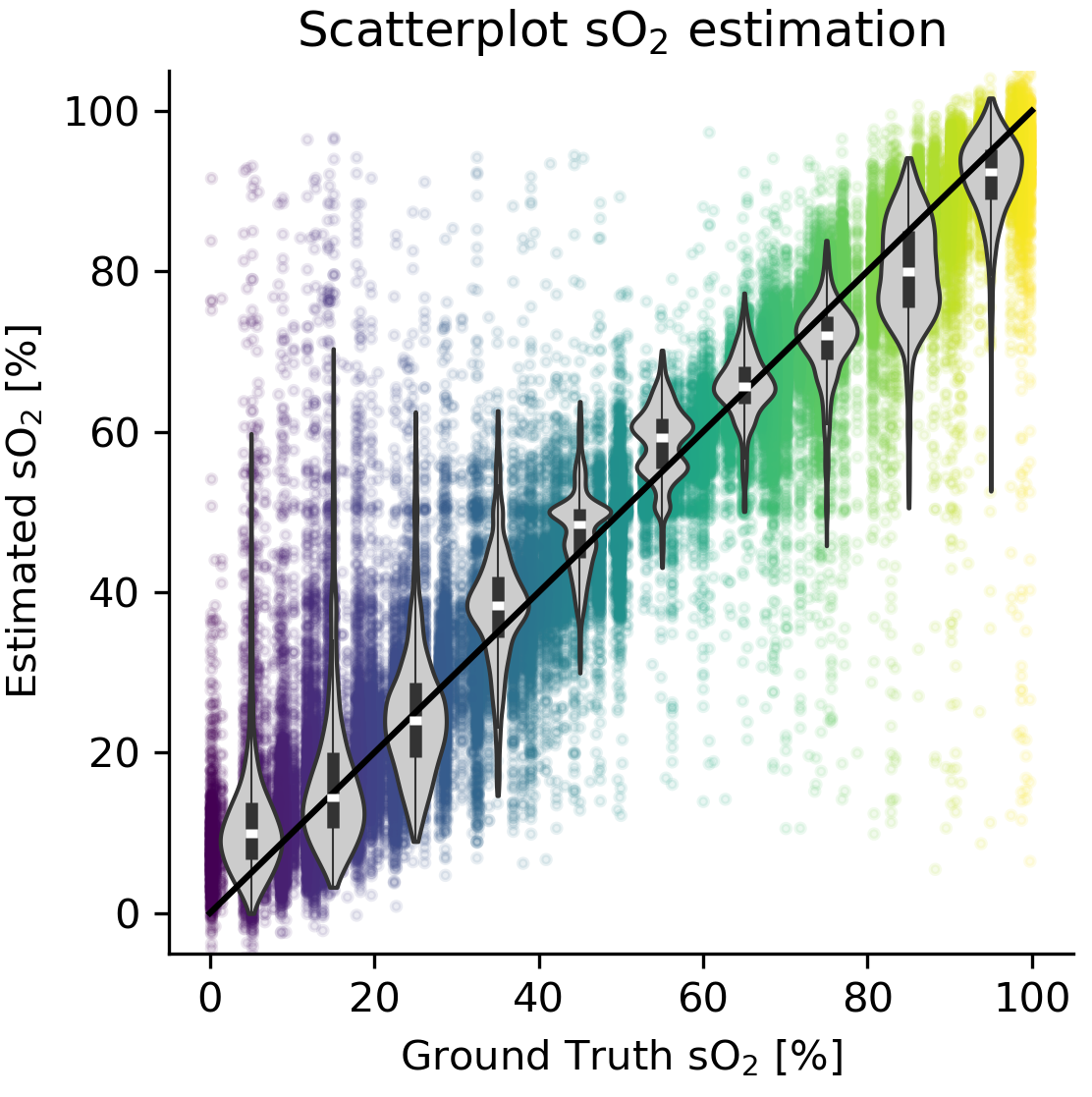}
\end{tabular}
\end{center}
\caption[example] 
{ \label{fig:res:insilico} 
Scatterplot of ground truth sO$_2$ and  $\hat{\textrm{sO}}_2$ estimated from the $\textrm{p}_0$ spectra with our algorithm. The bisector of the graph is indicated by a black line. In the violin plots, the black line represents the 25th and 75th percentile and the white box represents the median.}
\end{figure}

We use 80\% of the \emph{in silico} dataset for training, 10\% for supervision of the training process and hyperparameter optimization, and the remaining 10\% as a test set to report the \emph{in silico} results. We report the relative sO$_2$ estimation error e$_{\textrm{sO}_2} = (\hat{sO}_2 - \textrm{sO}_2) / \textrm{sO}_2$, where $\hat{\textrm{sO}}_2$ represents the estimated blood oxygen saturation and $\textrm{sO}_2$ represents the ground truth simulated blood oxygenation. All \emph{in silico} estimation results for the test data are shown in Figure \ref{fig:res:insilico}. Additionally, violin plots present the error distribution within each interval of ten percentage points (0\,--\,10\%, 10\,--\,20\%, etc). The median relative estimation error  e$_{\textrm{sO}_2}$ was 6.1\% with an interquartile range of (2.4\%,\,18.7\%).

\subsection*{Imaging of porcine brain}

As our first \textit{in vivo} experiment, we applied our method to images of a porcine brain during open surgery. These images were recorded at the same wavelengths as in the training dataset. We evaluate a single series of images which were normalized by the recorded laser energy and reconstructed with the delay-and-sum algorithm using a hamming window. For the calculation of the mean oxygenation within the ROI, only those point estimates were taken into consideration, where the signal at the isosbestic point of ~800\,nm was greater than a noise equivalent level.

\begin{figure} [ht]
\begin{center}
\begin{tabular}{c} 
\includegraphics[width=8cm]{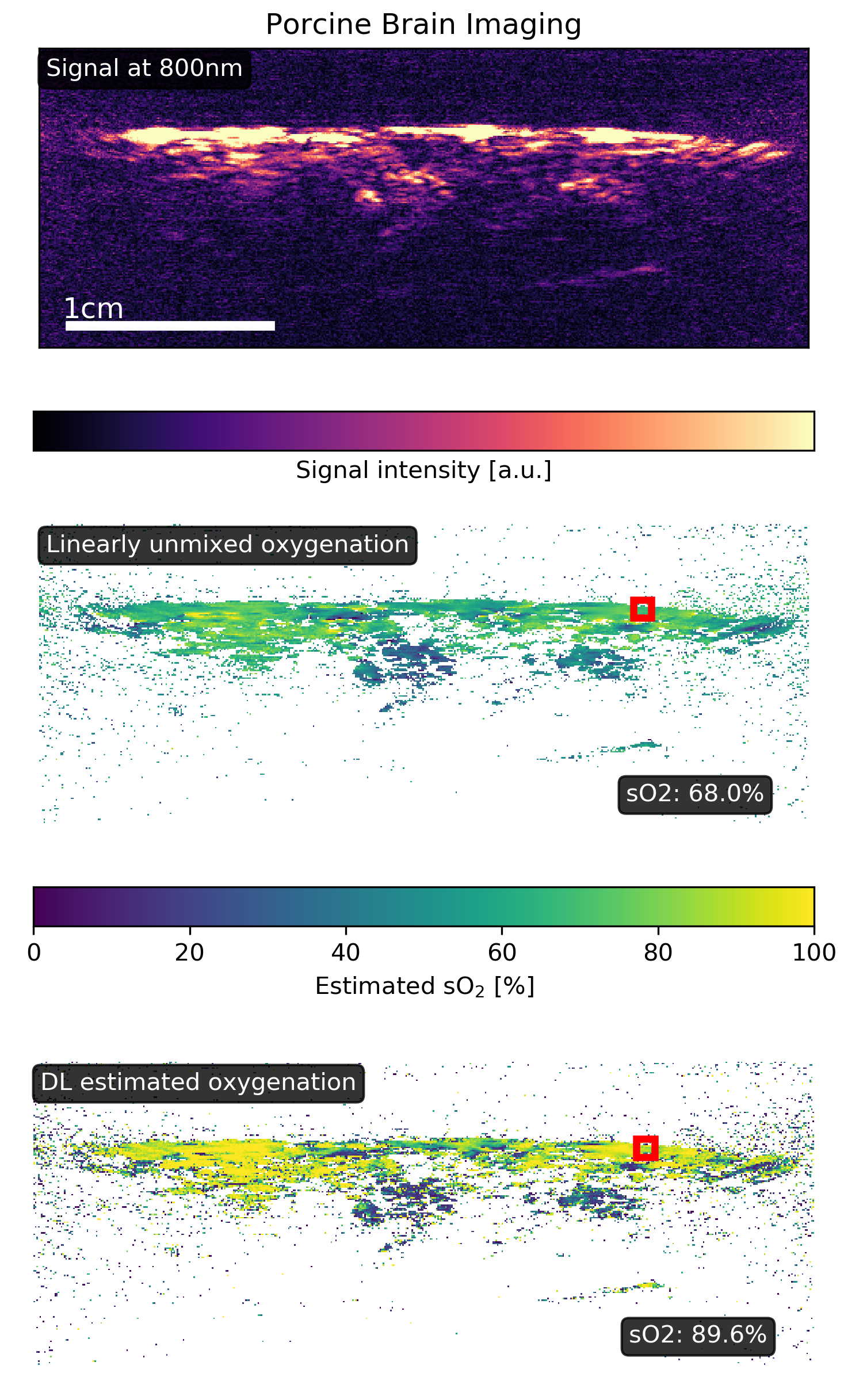}
\end{tabular}
\end{center}
\caption[example] 
{ \label{fig:res:invivo:brain} 
Qualitative example of sO$_2$ estimation derived from a multispectral PA image of a pig brain. The top image shows the signal at 800nm, the middle image shows the unmixing result using a linear spectral unmixing technique, and the bottom image shows the unmixing results with the proposed method. The red region of interest (ROI) shows a superficial region of highly oxygenated blood.}
\end{figure} 

In the image, the red bounding box marks an area where blood with a high oxygenation was present (cf. Figure \ref{fig:res:invivo:brain}). Our method estimated a blood oxygenation of about 90\% , whereas linear spectral unmixing using a QR decomposition yielded an estimate of 68\% in the same area.

\subsection*{Imaging of human forearm}

We also applied our method to \emph{in vivo} images of a the forearm of a healthy human volunteer also with the same 26 wavelengths as used in the simulations. This imaging scenario was specifically chosen as an out-of-distribution test for the method. This is, because no well matching spectra should be contained in our training set, as we did not consider the presence of melanin in our simulation. We normalized each image by the laser energy and reconstructed the images with the delay-and-sum algorithm using a hamming window. To decrease the inter frame variability, we averaged over 10 consecutive frames of the same wavelength. Prior to averaging we registered the images with an optical flow-based method to account for motion artifacts~\cite{kirchner2019open}.

\begin{figure} [ht]
\begin{center}
\begin{tabular}{c} 
\includegraphics[width=8cm]{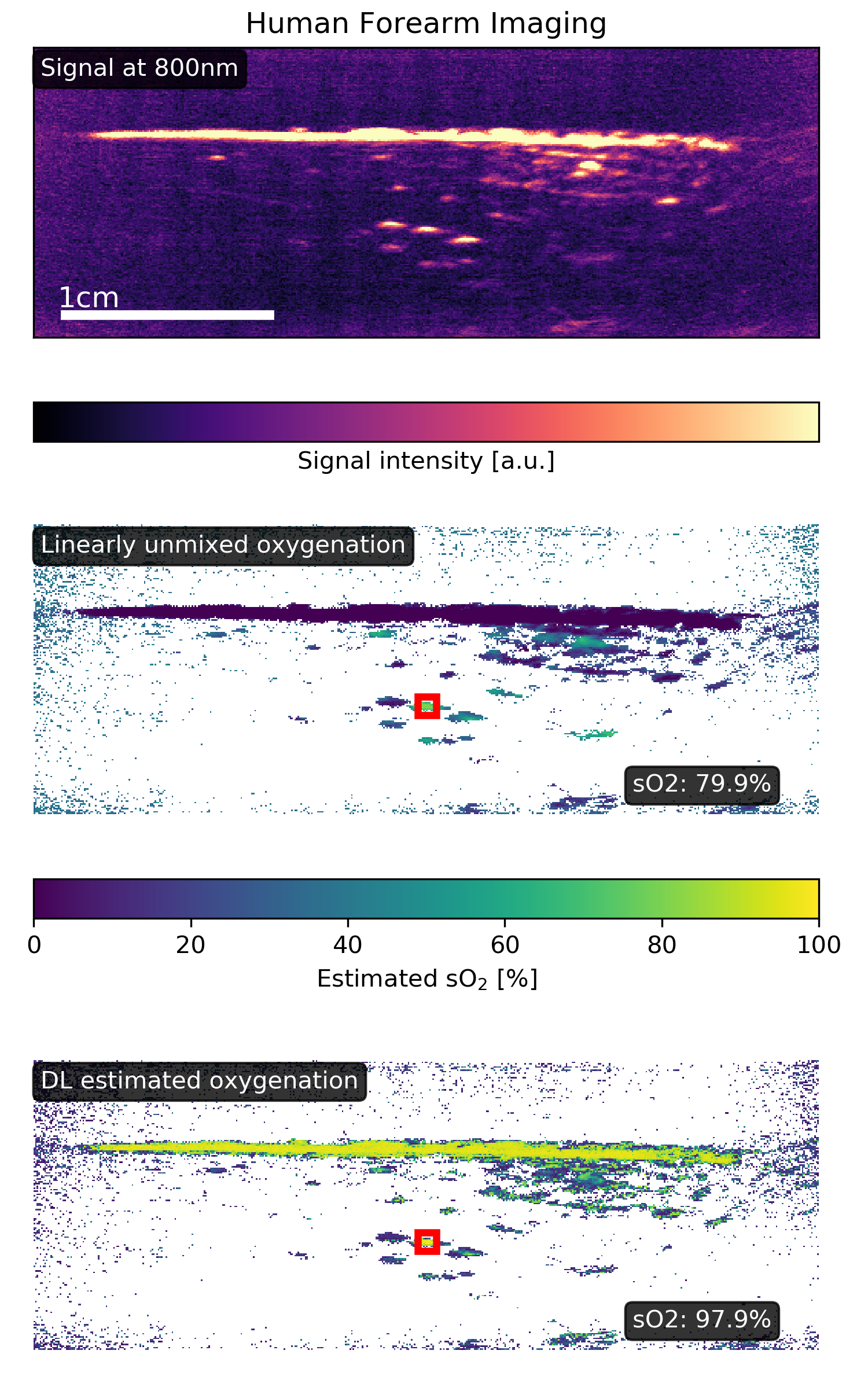}
\end{tabular}
\end{center}
\caption[example] 
{ \label{fig:res:invivo:forearm}
Qualitative example of sO$_2$ estimation derived from a multispectral PA image of the forearm of healthy human volunteers. The red ROI shows the radial artery at a depth of about five millimeters. The top image shows the signal at 800nm, the middle image shows the unmixing result using a linear spectral unmixing technique, and the bottom image shows the unmixing results with the proposed method.}
\end{figure}

In the oxygenation images shown in Figure \ref{fig:res:invivo:forearm}, the upper signal originating from the radial artery is chosen as the region of interest (ROI) and marked by a red bounding box. In this ROI, only those point estimates were taken into consideration for the averaged sO$_2$ result, where the signal at 800\,nm was greater than a noise equivalent level. The arterial blood oxygenation as determined by our algorithm was about 98\%, whereas spectral unmixing using a QR linear unmixing algorithm yielded a value of about 80\%.

\section{Discussion and Conclusion}
\label{sec:discussion}

In this work, we investigate the feasibility of inferring sO$_2$ from pixel-wise initial pressure spectra $S_{\textrm{p}_0}$ using a deep learning model trained on simulated data. While a photoacoustic signal is proportional to the product of optical absorption $\mu_a$ and the light fluence $\phi$, standard unmixing algorithms ignore the dependence on $\phi$, which can potentially result in large errors, especially deep in tissue. To overcome this issue, we propose the LSD-qPAI algorithm that learns a mapping function from pixel-wise multispectral initial pressure data to blood oxygenation. Our initial \emph{in silico} results look promising, yielding a median sO$_2$ estimation error of 6\% \textit{in silico}. The LSD-qPAI method is potentially relevant especially for clinical imaging with hand-held PA scanners, as sO$_2$ estimations with high accuracy and robustness can potentially be computed in real time. This has been demonstrated in similar diffuse reflectance multispectral imaging applications \cite{wirkert2016robust}.

In prior work, we achieved a comparable accuracy \textit{in silico} when estimating the optical absorption coefficient \cite{kirchner2018context, grohl2018confidence}. However, in contrast to these methods, we now are also able to obtain plausible results \emph{in vivo} when imaging an open pig brain. Here, the estimations of our method are closer to the expected arterial blood oxygenation values of healthy subjects (near 100\% \cite{Zander1990}). Especially in comparison to the results of classical linear unmixing, LSD-qPAI seems to provide physiologically more plausible estimations. 

Our method utilizes a normalization of the simulated and recorded spectra, sacrificing $\textrm{p}_0$ amplitude information to eliminate the need to calibrate the \emph{in silico} $\textrm{p}_0$ training data to the target domain. However, this also means that we discard one dimension of the feature vector, restricting the minimum amount of spectra needed to be able to reliably infer optical properties.

\begin{figure} [ht]
\begin{center}
\begin{tabular}{c} 
\includegraphics[width=8cm]{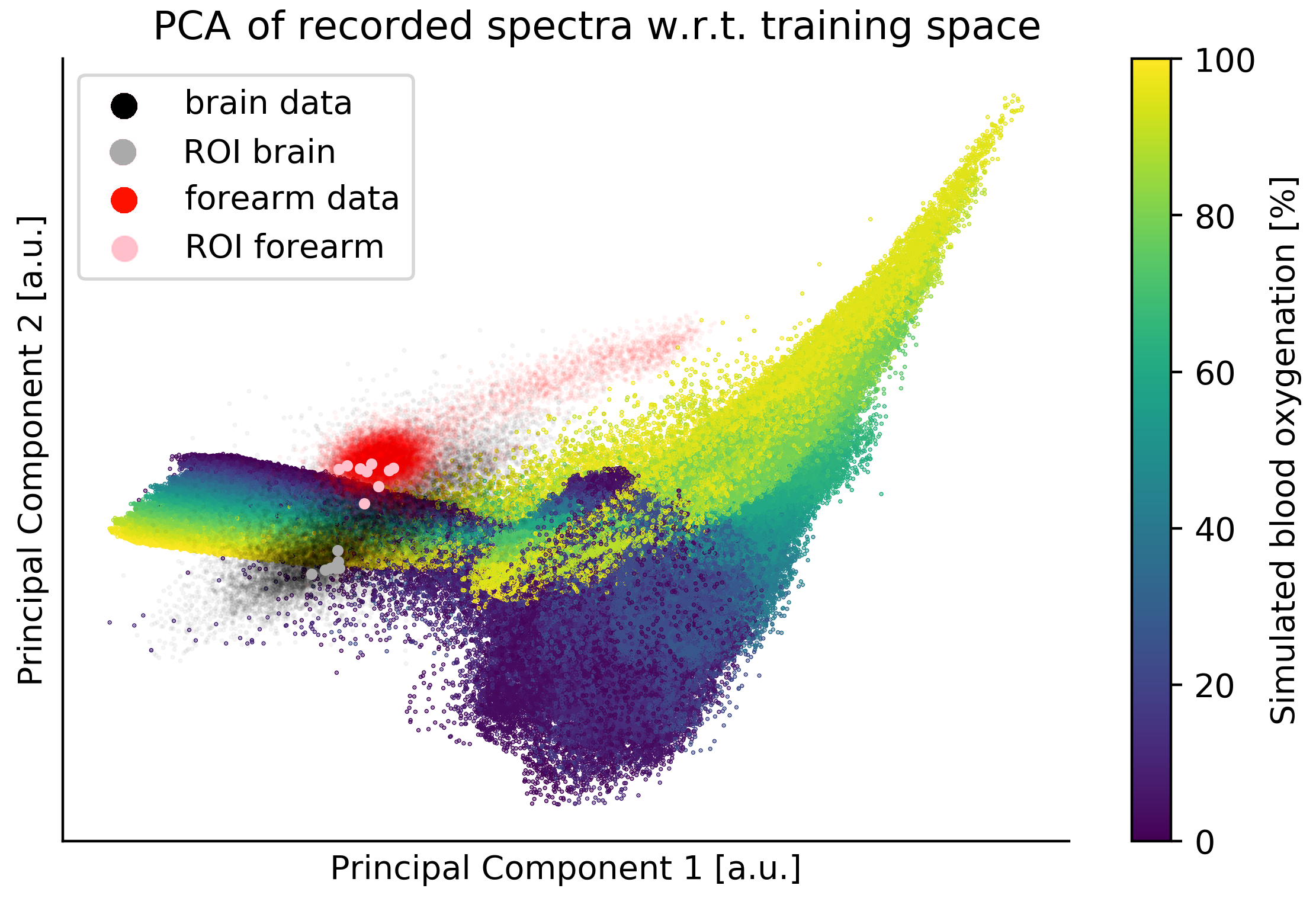}
\end{tabular}
\end{center}
\caption[example] 
{ \label{fig:dis:pca}
Principal component analysis (PCA) of the \emph{in vivo} recorded spectra of porcine brain tissue (black) and human forearm tissue (red). The samples of the regions of interest shown in Fig.\ \ref{fig:res:invivo:brain} and Fig.\ \ref{fig:res:invivo:forearm} and shown in gray and pink respectively. The principal component projection of the training data is also color-coded with the associated oxygenation values.}
\end{figure}

For the \emph{in vivo} experiments it was to be expected that the recorded \emph{in vivo} spectra of the brain images match our training distribution more closely compared to the forearm images. This is because light needs to penetrate through the skin in order to obtain images of the forearm, which contains melanin, a chromophore that was not included in the simulation framework. It can be seen that neither the proposed method nor linear unmixing can handle skin tissue well, estimating  implausible and even impossible sO$_2$ results. Projecting the \emph{in vivo} spectra to the first two principal components of the training data seems to confirm this hypothesis (cf.\ Fig.~\ref{fig:dis:pca}). In the figure it is apparent that most of the porcine spectra (colored in black) are contained within the support of the first two principal components of our simulation space, whereas this does not appear to be the case for the forearm spectra. In this context it should be noted that the first two principal components (computed on the training data) account for 95.4\% of the variation in our dataset. The projection image illustrates that the distribution of the training data match the distribution of the test (in vivo) data more closely when imaging pig brain instead of human forearm. 

When considering the large number of plausible tissue geometry and oxygenation configurations, the number of training samples used for this study was low. Also, no noise model was applied to the simulated spectra in addition to the simulation noise inherent to the Monte Carlo procedure. Hence, a thorough investigation of the influence of a realistic noise model on the spectra would be very interesting. Our a priori assumptions for the tissue parameters used in the simulation pipeline led to very difficult situations for the unmixing algorithm to resolve. For example, due to the 2\,\% blood volume fraction in the background, light fluence was the dominating factor even in shallow point absorbers - a behavior not usually observed in realistic scenarios, where less than 1\% would be a more realistic assumption~\cite{jacques2013optical}. Also, the simulation assumption that blood oxygenation is constant throughout the tissue is not generally correct and more realistic variations on these assumptions should be considered in future work.

Overall, the presented initial results \emph{in vivo} are very promising. However, a thorough and well-designed \emph{in vivo} validation needs to be performed to deduce meaningful conclusions regarding the general applicability of the presented method. Such future studies should cover a much broader range of possible sO$_2$ values and include more realistic scenarios when assessing the unmixing accuracy. Future work should also include a comprehensive comparison of this method to state-of-the-art methods, such as eigenspectra multispectral optoacoustic tomography (eMSOT) \cite{tzoumas2016eigenspectra}, the end-to-end qPAI method presented by Cai et al.~\cite{cai2018end}, as well as other linear and nonlinear spectral unmixing techniques \cite{keshava2002spectral,li2018photoacoustic}.

\section*{Acknowledgements}      
 
This project has received funding from the European Union’s Horizon 2020 research and innovation programme through the ERC starting grant {\footnotesize COMBIOSCOPY} under grant agreement No. ERC-2015-StG-37960. The authors would like to thank E. Santos, M. Herrera, and A. Hernández-Aguilera for provisioning of PA brain data, K. Dreher, N. Holzwarth, A. Klein, and S. Onogur for proof-reading the manuscript and the ITCF of the DKFZ for enabling extensive use of their computing cluster for data simulation.

\section*{Author Contributions}

Conceptualization, J.G., T.K., T.A., and L.M.-H.; Data curation, J.G. and T.K.; Formal analysis, J.G.; Funding acquisition, L.M.-H.; Investigation, T.K., T.A., and J.G.; Methodology, J.G. and T.K.; Project administration, L.M.-H.; Software, J.G.; Supervision, L.M.-H.; Validation, J.G. and T.K.; Visualization, J.G.; Method naming, T.K.; Writing—original draft, J.G.; and Writing—review and editing, J.G., T.K., T.A. and L.M.-H.



\begin{thebibliography}{10}

\bibitem{cox2009challenges}
Cox, B., Laufer, J., and Beard, P., ``The challenges for quantitative
  photoacoustic imaging,'' in [{\em Photons Plus Ultrasound: Imaging and
  Sensing 2009}{\nolinebreak\hspace{0.1em}]},   {\bf 7177},  717713,
  International Society for Optics and Photonics (2009).

\bibitem{wang2012photoacoustic}
Wang, L.~V. and Hu, S., ``Photoacoustic tomography: in vivo imaging from
  organelles to organs,'' {\em science}~{\bf 335}(6075),  1458--1462 (2012).

\bibitem{li2018photoacoustic}
Li, M., Tang, Y., and Yao, J., ``Photoacoustic tomography of blood oxygenation:
  A mini review,'' {\em Photoacoustics}  (2018).

\bibitem{maslov2007effects}
Maslov, K., Zhang, H.~F., and Wang, L.~V., ``Effects of wavelength-dependent
  fluence attenuation on the noninvasive photoacoustic imaging of hemoglobin
  oxygen saturation in subcutaneous vasculature in vivo,'' {\em Inverse
  Problems}~{\bf 23}(6),  S113 (2007).

\bibitem{bu2012model}
Bu, S., Liu, Z., Shiina, T., Kondo, K., Yamakawa, M., Fukutani, K., Someda, Y.,
  and Asao, Y., ``Model-based reconstruction integrated with fluence
  compensation for photoacoustic tomography,'' {\em IEEE Transactions on
  Biomedical Engineering}~{\bf 59}(5),  1354--1363 (2012).

\bibitem{zhao2017optical}
Zhao, L., Yang, M., Jiang, Y., and Li, C., ``Optical fluence compensation for
  handheld photoacoustic probe: An in vivo human study case,'' {\em Journal of
  Innovative Optical Health Sciences}~{\bf 10}(04),  1740002 (2017).

\bibitem{vogt2019photoacoustic}
Vogt, W.~C., Zhou, X., Andriani, R., Wear, K.~A., Pfefer, T.~J., and Garra,
  B.~S., ``Photoacoustic oximetry imaging performance evaluation using dynamic
  blood flow phantoms with tunable oxygen saturation,'' {\em Biomedical Optics
  Express}~{\bf 10}(2),  449--464 (2019).

\bibitem{cox2006two}
Cox, B.~T., Arridge, S.~R., K{\"o}stli, K.~P., and Beard, P.~C.,
  ``Two-dimensional quantitative photoacoustic image reconstruction of
  absorption distributions in scattering media by use of a simple iterative
  method,'' {\em Applied Optics}~{\bf 45}(8),  1866--1875 (2006).

\bibitem{laufer2006quantitative}
Laufer, J., Delpy, D., Elwell, C., and Beard, P., ``Quantitative spatially
  resolved measurement of tissue chromophore concentrations using photoacoustic
  spectroscopy: application to the measurement of blood oxygenation and
  haemoglobin concentration,'' {\em Physics in Medicine \& Biology}~{\bf
  52}(1),  141 (2006).

\bibitem{kirchner2018context}
Kirchner, T., Gr{\"o}hl, J., and Maier-Hein, L., ``Context encoding enables
  machine learning-based quantitative photoacoustics,'' {\em Journal of
  biomedical optics}~{\bf 23}(5),  056008 (2018).

\bibitem{cai2018end}
Cai, C., Deng, K., Ma, C., and Luo, J., ``End-to-end deep neural network for
  optical inversion in quantitative photoacoustic imaging,'' {\em Optics
  letters}~{\bf 43}(12),  2752--2755 (2018).

\bibitem{grohl2018confidence}
Gr{\"o}hl, J., Kirchner, T., Adler, T., and Maier-Hein, L., ``Confidence
  estimation for machine learning-based quantitative photoacoustics,'' {\em
  Journal of Imaging}~{\bf 4}(12),  147 (2018).

\bibitem{tzoumas2016eigenspectra}
Tzoumas, S., Nunes, A., Olefir, I., Stangl, S., Symvoulidis, P., Glasl, S.,
  Bayer, C., Multhoff, G., and Ntziachristos, V., ``Eigenspectra optoacoustic
  tomography achieves quantitative blood oxygenation imaging deep in tissues,''
  {\em Nature communications}~{\bf 7},  ncomms12121 (2016).

\bibitem{brochu2017towards}
Brochu, F.~M., Brunker, J., Joseph, J., Tomaszewski, M.~R., Morscher, S., and
  Bohndiek, S.~E., ``Towards quantitative evaluation of tissue absorption
  coefficients using light fluence correction in optoacoustic tomography,''
  {\em IEEE transactions on medical imaging}~{\bf 36}(1),  322--331 (2017).

\bibitem{tzoumas2014unmixing}
Tzoumas, S., Deliolanis, N., Morscher, S., and Ntziachristos, V., ``Unmixing
  molecular agents from absorbing tissue in multispectral optoacoustic
  tomography,'' {\em IEEE transactions on medical imaging}~{\bf 33}(1),  48--60
  (2014).

\bibitem{jacques2014coupling}
Jacques, S.~L., ``Coupling 3d monte carlo light transport in optically
  heterogeneous tissues to photoacoustic signal generation,'' {\em
  Photoacoustics}~{\bf 2}(4),  137--142 (2014).

\bibitem{paszke2017automatic}
Paszke, A., Gross, S., Chintala, S., Chanan, G., Yang, E., DeVito, Z., Lin, Z.,
  Desmaison, A., Antiga, L., and Lerer, A., ``Automatic differentiation in
  pytorch,'' (2017).

\bibitem{zimmerer_david_mic_dkfz_trixi_2018}
Zimmerer, D., Petersen, J., Koehler, G., Wasserthal, J., Adler, T., and
  Wirkert, A., ``{MIC}-{DKFZ}/trixi,'' (2018).

\bibitem{eigenweb}
Guennebaud, G., Jacob, B., et~al., ``Eigen v3.'' http://eigen.tuxfamily.org
  (2010).

\bibitem{jacques2013optical}
Jacques, S.~L., ``Optical properties of biological tissues: a review,'' {\em
  Physics in Medicine \& Biology}~{\bf 58}(11),  R37 (2013).

\bibitem{kirchner2019open}
Kirchner, T., Gr{\"o}hl, J., Sattler, F., Bischoff, M.~S., Laha, A., Nolden,
  M., and Maier-Hein, L., ``An open-source software platform for translational
  photoacoustic research and its application to motion-corrected blood
  oxygenation estimation,'' {\em arXiv preprint arXiv:1901.09781}  (2019).

\bibitem{wirkert2016robust}
Wirkert, S.~J., Kenngott, H., Mayer, B., Mietkowski, P., Wagner, M., Sauer, P.,
  Clancy, N.~T., Elson, D.~S., and Maier-Hein, L., ``Robust near real-time
  estimation of physiological parameters from megapixel multispectral images
  with inverse monte carlo and random forest regression,'' {\em International
  journal of computer assisted radiology and surgery}~{\bf 11}(6),  909--917
  (2016).

\bibitem{Zander1990}
Zander, R., ``The oxygen status of arterial human blood,'' {\em Scandinavian
  Journal of Clinical and Laboratory Investigation}~{\bf 50}(sup203),  187--196
  (1990).

\bibitem{keshava2002spectral}
Keshava, N. and Mustard, J.~F., ``Spectral unmixing,'' {\em IEEE signal
  processing magazine}~{\bf 19}(1),  44--57 (2002).

\end{thebibliography}


\section*{Bibliography}
\end{document}